\documentclass[12pt,superscriptaddress,preprint,nofootinbib]{revtex4-1}
\usepackage{amssymb,graphics,graphicx,tabularx,color,epstopdf}
\usepackage{psfrag}
\usepackage{physics} % includes amsmath, plus many physics typesetting commands
\usepackage{amssymb}
\usepackage{amsmath}
\usepackage{bm} %bold math symbols
\usepackage{graphicx}
\usepackage{wrapfig}
\usepackage{natbib}
\usepackage[colorlinks=true,urlcolor=blue,bookmarks=true,citecolor=blue,breaklinks=true,linkcolor=black]{hyperref}
\usepackage{epstopdf}
\usepackage{enumerate}
\usepackage{xcolor}
\usepackage{float}
\usepackage{longtable}
\usepackage{gensymb}
\usepackage{braket}
\graphicspath{{images/}}
\def\2tvec#1#2{
\left(
\begin{array}{c}
#1  \\
#2  \\   
\end{array}
\right)}

\def\3tvec#1#2#3{
\left(
\begin{array}{c}
#1  \\
#2  \\   
#3  \\
\end{array}
\right)}

\begin{document}
\title{A Gauged Flavor Model of Quarks and Leptons}
\author{Bradley L. Rachlin\footnote{bradleyrachlin@gmail.com}$^{1}$ and Thomas W. Kephart\footnote{tom.kephart@gmail.com}}
\affiliation{  Department of Physics and Astronomy, Vanderbilt University, Nashville, Tennessee 37235, USA}
\date{\today}

\begin{abstract}
Beyond Standard Model physics frequently connects
flavor symmetry with a discrete group. If
the discrete symmetry arises spontaneously  from a gauge theory, one can maintain compatibility with
quantum gravity and avoid anomalies. We provide an example of such a model with the Standard Model gauge group extended to $SU(3)_C  \times SU(2)_L \times U(1)_Y \times SU(2)_{T'} $ where the binary tetrahedral flavor group
$T'$ is embedded in  $SU(2)_{T'}$. Quark and lepton masses and mixing angles are fit to data, where lepton mixing angles are shifted from tribimaximal values by the addition of scalar VEVs to agree with the experimental data.\\

%\vspace{5cm}

%\begin{center}
%{\color[rgb]{1,0.5,0} {\Huge DRAFT}}
%\end{center}
\end{abstract}
\maketitle

%\begin{document}
\title{SU(2) $\rightarrow$ $T^{'}$ Model for
Adjusted Tribimaximal Mixing and Cabibbo Angle}
\author{Bradley L. Rachlin\footnote{bradleyrachlin@gmail.com}$^{1}$ and Thomas W. Kephart\footnote{tom.kephart@gmail.com}}
\affiliation{  Department of Physics and Astronomy, Vanderbilt University, Nashville, Tennessee 37235, USA}
\date{\today}

\bigskip
\section{Introduction}
Flavor models of elementary particles have had to evolve as new data becomes available. As the data becomes more precise, the models become more sophisticated. The usual model building practice is to extend the standard model (SM) with a discrete symmetry which is used to fit the data. But variations abound, from extending a supersymmetric SM, to discrete group extended grand unified models (For reviews see \cite{Frampton:1994rk,Ishimori:2010au,Altarelli:2010gt,King:2013eh,King:2014nza}), to top-down fully gauged theories where the gauge group is sufficiently large to accommodate both the GUT and flavor symmetries \cite{Albright:2016lpi}. Here we take a minimalist approach and look for the smallest fully gauged model that can explain all the data. 

One of the simplest and most natural flavor models is the SM extended by the discrete group $T'$ \cite{Frampton:1994rk,Aranda:1999kc,Chen:2007afa,Frampton:2007et,Frampton:2008bz,Frampton:2010uw,Natale:2016xob,Carone:2016xsi}, where the one and two dimensional irreducible representations (irreps) accommodate the quarks,  while the leptons fit naturally into one and three dimensional irreps. For a phenomenological discussion and recent summary of the data, see e.g., \cite{Esteban:2016qun}. The current challenge is to fit the most recent neutrino data with a $T'$ model. A shortcoming of nearly all discrete flavor models is their lack of compliance with gravity \cite{Krauss:1988zc}, i.e., gravity breaks discrete global symmetries. But since gravity does not interfere with gauge symmetries, gauging a discrete symmetry by embedding it in a gauge group is a way to avoid this problem. But one still has to contend with discrete \cite{Ibanez:1991pr,Ibanez:1991hv,Luhn:2008sa,Fallbacher:2015pga,Talbert:2018nkq} or continuous chiral gauge anomalies.  Our minimalist approach then leads us to gauge $T'$ flavor. The smallest continuous group that contains $T'$ is 
$SU(2)$, so this is what we will attempt below. Various complications arise, but we will be able to deal with them as we go along. 

We take the simplified renormalizable $T'$ extension of the standard model of   \cite{Frampton:2008bz} and augment it in two ways. First, we add   scalar singlets, that will acquire VEVs and shift the predictions of tribimaximal (TBM) mixing and of the Cabibbo angle from  previous models   to be more in line with current experimental values. Second, after adding a few fermions, the $T'$ group is embedded into a gauged $SU(2)$ group we will call $SU(2)_{T'}$. This   averts problems with gravity and chiral anomalies that can arise from adding discrete groups to the standard model. It also provides an elegant description of the discrete symmetry as a residue of a gauge group acting at higher scale. 
%The standard model (SM) leptons fit into its $SU(2)$ extension and is otherwise unchanged, but the quarksector needs modification as described below.
%The extra scalars needed to shift the $T'$ predictions for the mixing angles conveniently fit in representations of $SU(2)$.
%WE DIDN'T END UP DETAILING THE SU2 IRREPS OF THE SCALARS
Finally we summarize how $SU(2)_{T'}$ can be broken directly to $T'$ with a VEV for a particular scalar multiplet.

The next section contains the lepton sector particle assignments, plus the assignments for the scalar fields that enter the lepton Yukawa Lagrangian at the $T'$ scale. Section III contains similar information for the quark sector; in Section IV we discuss tribimaximal (TBM) mixing, where a $T'$ triplet Higgs gets a vacuum expectation value (VEV). Since there is currently tension between the data and TBM predictions, we add $T'$ scalar singlets with VEVs to shift TBM predictions in Section V, where we show our new fit is in agreement with all lepton data. Section VI focuses on the quark sector, where the new scalar singlet VEVs now contribute to quark mixing. 

It is the above described $T'$ model we gauge to $SU(2)_{T'}$, and describe in Section VII, where various additional particles need to be added to avoid all chiral anomalies. Section VIII describes the spontaneous symmetry breaking (SSB) from $SU(2)_{T'}$ to $T'$, and Section IX contains our conclusions and plans for further work. An appendix collects all the $T'$ group theory needed for this paper.

\section{Lepton Sector Lagrangian at the $T'$ Scale}
We begin by reviewing the lepton sector just above the $T'$ scale. Because none of the  leptons will be in even dimensional irreducible representations (irreps), this sector is equivalent to an $A_4$ model \cite{Ma:2001dn,Babu:2002dz}. We have also given the model a $Z_2$ symmetry in order to disallow certain terms in the Lagrangian. This $Z_2$ will also be gauged.

The standard model leptons are assigned to the following irreps \cite{Frampton:2008bz} of $T'\times Z_2$ (and of $A_4 \times Z_2$):
\begin{equation}
\label{leptonreps}
\begin{array}{ccc}
\left. \begin{array}{c}
\left( \begin{array}{c} \nu_{\tau} \\ \tau^- \end{array} \right)_{L} \\
\left( \begin{array}{c} \nu_{\mu} \\ \mu^- \end{array} \right)_{L} \\
\left( \begin{array}{c} \nu_{e} \\ e^- \end{array} \right)_{L} 
\end{array} \right\} 
L_L  ({\bf 3},0)  &
\begin{array}{c}
~ \tau^-_{R}~ ({\bf 1_1},1)   \\
~ \mu^-_{R} ~ ({\bf 1_2},1) \\
~ e^-_{R} ~ ({\bf 1_3},1)  \end{array}
&
\left. \begin{array}{c}
~ N^{(1)}_{R}  \\
~ N^{(2)}_R  \\
~ N^{(3)}_{R} \\  \end{array} \right\}
N_R ({\bf 3},0)
\end{array}
\end{equation}
where $N_R$ is a $T'$ triplet of right handed neutrinos. In addition we will need the following scalars:
\begin{equation}
\label{scalarreps0}
\begin{array}{c}
H_3({\bf 3},0)=(H_3^1,H_3^2,H_3^3)\\
H_3'({\bf 3},1)=(H_{3'}^1,H_{3'}^2,H_{3'}^3)\\
H_{1_1}({\bf 1_1},0)\\
H_{1_2}({\bf 1_2},0)\\
H_{1_3}({\bf 1_3},0)
\end{array}
\end{equation}
Where the subscripts correspond to the $T'$ irrep where the scalars live.

Aside: Note that here and and below we use a different notation from \cite{Frampton:2008bz} which used a multiplicative
form for the  $Z_2$ charges, i.e., $\pm 1$. Since we will be concerned with discrete and continuous chiral gauge anomalies,
we use additive  $Z_2$ charges, i.e., integers mod 2, to be consistent with most of the literature. When we later embed  $Z_2$ in a $U(1)$ we will use integer charges.

With the above content, the most general lepton sector Yukawa Lagrangian is:
\begin{equation}\label{eq:leptonL}
{\cal L}_{l}= Y_\tau L_L \tau_R H_{3'} +Y_\mu L_L \mu_R H_{3'}+ Y_e L_L e_R H_{3'}+ L_L N_R (Y_x H_{1_2}+Y_y H_{1_3}+Y_z H_{1_1}+Y_{TB} H_3)+ m_N N_R N_R
\end{equation}
The proper choice of VEVs for $H_3$ and $H_{3'}$ lead to values for the charged masses and the TBM mixing matrix. Giving VEVs to the singlets will break $T'$ to $Q$, the  group of unit quaternions, and shift the TBM matrix closer to experimentally compatible values.

\section{Quark sector Lagrangian at the $T'$ Scale}
The main advantage of a $T'$ flavor model is that it is the  discrete group of smallest order with a sufficiently diverse set of irreps that can be used to model both the quark and lepton sectors. Specifically,  it has even-dimensional irreps that can also be used to economically describe the quark sector, as we will now summarize \cite{Frampton:2008bz}. The standard model quarks are assigned to the following irreps:
\begin{equation}
\label{quarkreps}
\begin{array}{cc}
\left( \begin{array}{c} t \\ b \end{array} \right)_{L}
~ {\cal Q}_L ~~~~~~~~~~~ ({\bf 1_1},0)   \\
\left. \begin{array}{c} \left( \begin{array}{c} c \\ s \end{array} \right)_{L}
\\
\left( \begin{array}{c} u \\ d \end{array} \right)_{L}  \end{array} \right\}
Q_L ~~~~~~~~ ({\bf 2_1},0)
\end{array}
\begin{array}{c}
t_{R} ~~~~~~~~~~~~~~ ({\bf 1_1},0)   \\
b_{R} ~~~~~~~~~~~~~~ ({\bf 1_2},1)  \\
\left. \begin{array}{c} c_{R} \\ u_{R} \end{array} \right\}
{\cal C}_R ~~~~~~~~ ({\bf 2_2},1)\\
\left. \begin{array}{c} s_{R} \\ d_{R} \end{array} \right\}
{\cal S}_R ~~~~~~~~ ({\bf 2_3},0).
\end{array}
\end{equation}

In addition to the scalars listed above, we  add one more singlet:
\begin{equation}
\label{scalarreps}
\begin{array}{c}
%H_{1_1}({\bf 1_1},+1)\\
%H_{1_2'}({\bf 1_2},1)\\
H_{1_3'}({\bf 1_3},1)
\end{array}
\end{equation}
Hence the most general quark sector Yukawa Lagrangian is then:
\begin{equation}\label{eq:quarkLag}
{\cal L}_{q}= Y_t{\cal Q}_L t_R H_{1_1}+ Y_b{\cal Q}_L b_R H_{1_{3'}}+ Q_L {\cal C}_R (Y_{\cal C}H_{3'}+Y_{\cal C'} H_{1_{3'}})+Q_L {\cal S}_R(Y_{\cal S}H_3 +Y_{\cal S'}H_{1_2})
\end{equation}

We see that a constraint on our model is that the VEVs of $H_3,H_{3'},H_{1_2},\text{and} ~ H_{1_1}$ must have values that are simultaneously compatible with the experiment data for both the quark and lepton sector.

\section{TBM Mixing from $T'$}

Before we derive our experimentally compatible PMNS matrix  \cite{PMNS}, we show that just below  the $T'$ energy scale where only $T'$ triplets have VEVs, the neutrinos exhibit the familiar TBM mixing pattern \cite{Harrison:2002er}. Using the Clebsch-Gordan coefficients for $T'$ detailed in the Appendix, we find that the term $m_N N_R N_R$ from equation (\ref{eq:leptonL}) gives a mass matrix for right handed neutrinos:
\begin{equation}
M_N=
\begin{pmatrix}
m_N&0&0\\
0&0&m_N\\
0&m_N&0
\end{pmatrix}
\end{equation}

Similarly, we construct the Dirac mass matrix associated with the term $Y_{TB} L_L N_R H_3$ of the lepton Lagrangian:
\begin{equation}
M_D=\begin{pmatrix}
eN_1&eN_2&eN_3\\
\mu N_1&\mu N_2 &\mu N_3\\
\tau N_1&\tau N_2 &\tau N_3
\end{pmatrix}=Y_{TB}
\begin{pmatrix}
v_2&-v_1&0\\
-v_3&0&v_1\\
0&v_3&-v_2
\end{pmatrix}
\end{equation}

Where $(v_1,v_2,v_3)$ is the VEV of the scalar $H_3$. The Majorana mass matrix is given by:
\begin{equation}
M_\nu=M_DM_N^{-1}M_D^T
\end{equation}

The rows of the Majorana mixing matrix are the normalized eigenvectors of this mass matrix, we find that for a VEV of $<H_3> = V(1,1,-2)$, (where V is some constant), we recover the TB mixing matrix in the form
\begin{equation}
U_{TBM}=
\begin{pmatrix}
\sqrt{\frac{2}{3}}&\sqrt{\frac{1}{3}}&0\\
-\sqrt{\frac{1}{6}}&\sqrt{\frac{1}{3}}&-\sqrt{\frac{1}{2}}\\
-\sqrt{\frac{1}{6}}&\sqrt{\frac{1}{3}}&\sqrt{\frac{1}{2}}
\end{pmatrix}.
\end{equation}
Currently TBM is excluded at the $5\sigma$ level. For a different perspective see \cite{Rahat:2018sgs}.

\section{Shifted TBM Mixing}

Our next step is to augment this matrix using VEVs for the additional  scalars $H_{1_1}$, $H_{1_2}$ and $H_{1_3}$. 
(For an alternative  perturbation theory  approach see \cite{Eby:2008uc}.) Including these in the model introduces the terms $L_L N_R (Y_xH_{1_2}+Y_yH_{1_3}+Y_zH_{1_1})$ into the Lagrangian. These terms have a mass matrix:
\begin{equation}
M_{xyz}=
\begin{pmatrix}
-x&z&y\\
-y&x&z\\
z&-y&-x
\end{pmatrix}
\end{equation}
Where $x$, $y$, and $z$ represent $Y_x<H_{1_2}>$, $Y_y<H_{1_3}>$, and $Y_z<H_{1_1}>$ respectively. Our Dirac mass matrix is now
\begin{equation}
M_{D'}=M_D+M_{xyz}=
Y_{TB}
\begin{pmatrix}
v_2-x'&-v_1+z'&y'\\
-v_3-y'&x'&v_1+z'\\
z'&v_3-y'&-v_2-x'
\end{pmatrix}
\end{equation}
where $x'=\frac{x}{Y_{TB}}$, $y'=\frac{y}{Y_{TB}}$ and $z'=\frac{z}{Y_{TB}}$. The Majorana mixing matrix, U, is obtained the same way as before. The fit parameters $x'$, $y'$ and $z'$ can now be varied   to shift the entries of U from their TBM mixing values closer to current experimental values. The present 3$\sigma$ experimental ranges of the magnitudes of the matrix elements are given below \cite{Esteban:2016qun,NuFIT2018,Tanabashi:2018}:
\begin{equation}
\begin{pmatrix}
0.799 \leftrightarrow 0.844 &\quad  0.516 \leftrightarrow 0.582 &\quad 0.141  \leftrightarrow 0.156\\
0.242 \leftrightarrow 0.494 &  \quad 0.467 \leftrightarrow 0.678&  \quad 0.639 \leftrightarrow 0.774\\
0.284\leftrightarrow 0.521 & \quad 0.490 \leftrightarrow 0.695 &\quad 0.615 \leftrightarrow 0.754 \\
\end{pmatrix}
\end{equation}

\noindent The next step we can take is to vary the parameters $x'$, $y'$, and $z'$ from -1 to 1 (within a reasonable precision), and find the values for which the least accurate elements' error is minimized. We find that to the nearest hundreth, this minimum is obtained at $(x',y',z')=(0.32,-0.26,-.40)$ with maximum error of 1.922$\sigma$. Explicitly, these values correspond to a mixing matrix:
\begin{equation}
\begin{pmatrix}
-0.829&0.539&0.148\\
0.289&0.640&-0.712\\
0.478&0.548&0.686
\end{pmatrix}
\end{equation}
which can be compared to the experimental numbers in eq. (13).

The errors relative to experiment are given below in units of $\sigma$:
\begin{equation}
\begin{pmatrix}
1.029&0.910&0.228\\
1.882&1.919&0.224\\
1.922&1.313&0.083
\end{pmatrix}.
\end{equation}

In addition to minimizing the error of the least accurate entry we can minimize the average error of the matrix elements.

Looping over all possible values of $x'$, $y'$, and $z'$ (again to the nearest hundredth) minimizes the mean error at $(0.32,-0.27,-.045)$, with a value of 0.870 $\sigma$. Our mixing matrix is now
\begin{equation}
\begin{pmatrix}
-0.822&0.549&0.149\\
0.298&0.638&-0.710\\
0.485&0.539&0.688
\end{pmatrix},
\end{equation}

with errors

\begin{equation}
\begin{pmatrix}
0.090&0.0351&0.199\\
1.671&1.869&0.151\\
2.091&1.558&0.167
\end{pmatrix}.
\end{equation}

From both these perspectives on error analysis, our $T'$ model extended with a pair of scalar singlets agrees with the current experimental data which provides a significant improvement over the simple TBM model with singlet VEVs on the same order as the triplet VEVs, i.e., without introducing a new length scale.

We can also examine our fitting with a contour plot. Our error values are the most sensitive to changes in $x'$ so we hold it constant at 0.32 and allow our parameters $y'$ and $z'$ to vary between -1 and 0 as shown in the plots below:
(The parameter range $(y', z')>0$ gives very high error values so it is not shown.) 
\begin{figure}[H]

\includegraphics[scale=1]{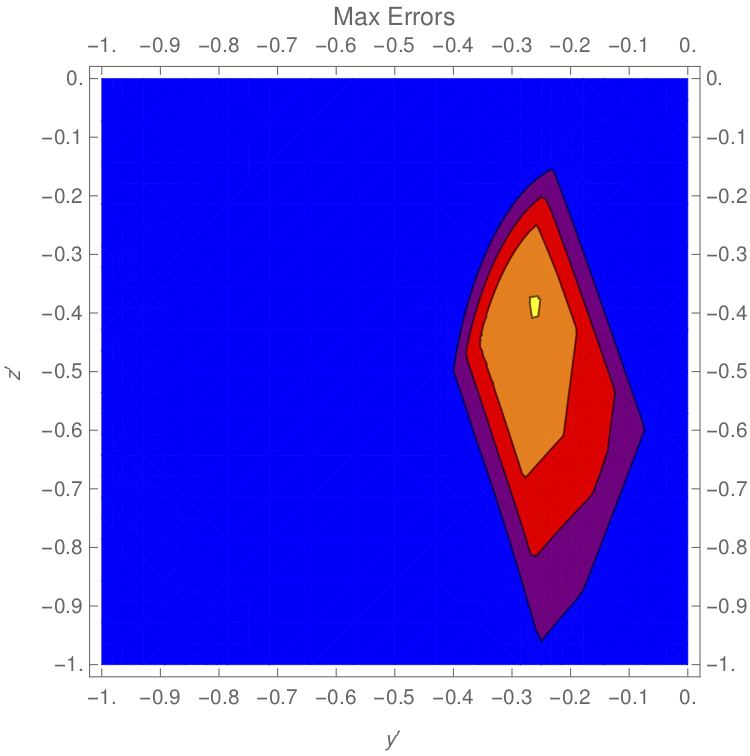}
\includegraphics[scale=.7]{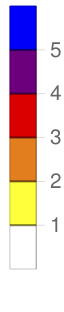}
\centering
\caption{Contour plot with $x'$ fixed at 0.32 of maximum mixing matrix error relative to experimental data, where values are in units of $\sigma$.}
\end{figure}

\begin{figure}[H]
\includegraphics[scale=1]{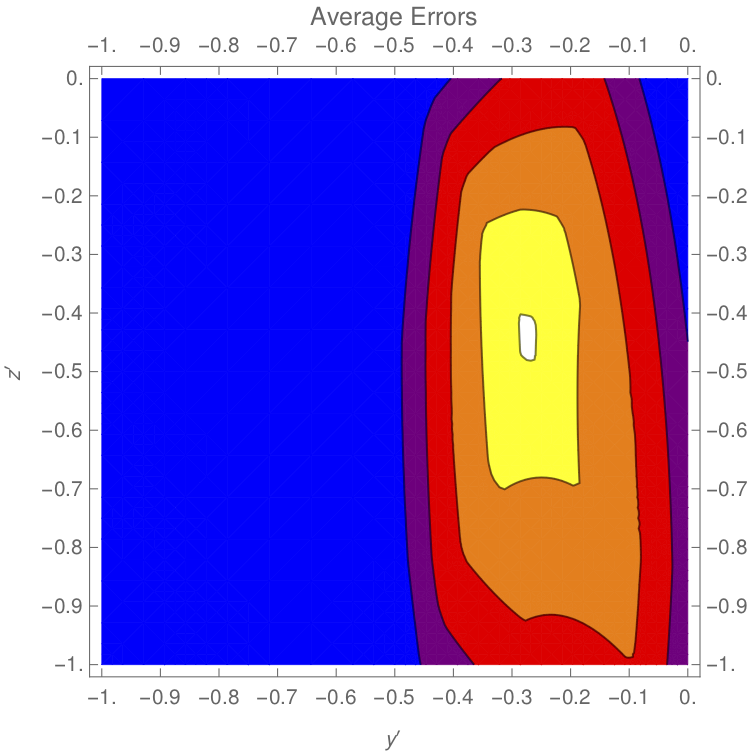}
\includegraphics[scale=.7]{contourlegend.png}
\centering
\caption{Contour plot with $x'$ fixed at 0.32 of average mixing matrix error relative to experimental data, where values are in units of $\sigma$.}
\end{figure}

We see from these plots that there are relatively small ranges, but still without fine tuning beyond an order of magnitude,  for our parameters that give us maximum error and average error less that $2\sigma$ and $1\sigma$ respectively.

\section{Quark Mixing}

As shown in \cite{Frampton:2008bz}, one can derive a reasonable prediction for the Cabbibo angle from the Lagrangian in equation (\ref{eq:quarkLag}). We rederive this result here for our basis and then augment the value when an additional scalar  has a VEV. We also find the mass matrices for the first two generations of up and down type quarks from the terms $Q_L {\cal C}_R H_{3'}$ and $Q_L {\cal S}_R H_{3}$ respectively. From the discussion above we know that $H_3$ must have a VEV of the form $V(1,1,-2)$. To give the correct masses to the charged leptons $H_{3'}$ must have a VEV:
\begin{equation}
<H_{3'}> = \begin{pmatrix}
\frac{m_{\tau}}{Y_{\tau}},\frac{m_{\mu}}{Y_{\mu}},\frac{m_{e}}{Y_{e}}
\end{pmatrix}
\end{equation}.

Using these values along with the Clebsch-Gordan coefficients found in the Appendix, we obtain a mass matrix $U$ for the up-type quarks:
\begin{equation}
U=Y_{\cal C}
\begin{pmatrix}
\frac{m_{\mu}}{Y_{\mu}}&-\frac{1}{\sqrt{2}}\frac{m_{e}}{Y_{e}}\\
-\frac{1}{\sqrt{2}}\frac{m_{e}}{Y_{e}}& \frac{m_{\tau}}{Y_{\tau}}
\end{pmatrix}
\end{equation}.

Because $m_{\tau}>m_{\mu}>>m_e$ we can, to lowest order,  set $m_e=0$ which gives a diagonal mass matrix U. For our down-type mass matrix, $D$ we obtain:
\begin{equation}
D=Y_{\cal S}
\begin{pmatrix}
-1&-\frac{1}{\sqrt{2}}\\
-\frac{1}{\sqrt{2}}&2
\end{pmatrix}
\end{equation}.

The mixing matrix for the first two quark generations, (the upper left corner of the CKM matrix) is $W = K_u^\dagger K_d$. Where $K_u$ and $K_d$ are the unitary matrices that diagonalize the Hermitian squares of $U$ and $D$ respectively. Since $U$ is already diagonal, $K_u$ is just the identity matrix and we have:
\begin{equation}
DD^\dagger=Y_{\cal S}^2
\begin{pmatrix}
\frac{3}{2}&-\frac{1}{\sqrt{2}}\\
-\frac{1}{\sqrt{2}}&\frac{9}{2}
\end{pmatrix}
\end{equation}

\begin{equation}
W= K_d =
\begin{pmatrix}
0.976&0.218\\
-0.218&0.976
\end{pmatrix}
\end{equation}

This gives an expression for the Cabibbo angle $\Theta$
\begin{equation}
\text{tan}(2\Theta)=\begin{pmatrix}\frac{\sqrt{2}}{3}\end{pmatrix}~\rightarrow ~\text{sin}(\Theta)=.218
\end{equation}
in agreement with \cite{Frampton:2008bz}.

We can see that this prediction roughly approximates the measured value of $sin\Theta \approx .225$. But given the high experimental precision of the Cabibbo angle, this prediction disagrees with experiment by over twelve standard deviations. Some of this variation can be explained by the fact that we have not included mixing with the third family in this simplified model. 
We could include third family mixing by adding the terms to the Lagrangian in equation (\ref{eq:quarkLag}):
\begin{equation}
{\cal L}'_{q}= Y_{tcu}{\cal Q}_L {\cal C}_R H_{2_3} + Y_{bsd}{\cal Q}_L {\cal S}_R H_{2_2}+Y_{cut}Q_L t_R H_{2_1}+Y_{sdb}Q_L b_R H_{2_3},
\end{equation}
but this introduces at least six more free parameters into the theory, and since we know such contributions to be very small we ignore them in the present analysis. Instead we can shift our prediction to well within the one sigma range by giving a VEV to the scalar $H_{1_3}$ in the quark Lagrangian in eq.(\ref{eq:quarkLag}). With this additional term, the down-type quark matrix becomes:
\begin{equation}
D'=Y_{\cal S}
\begin{pmatrix}
-1&\frac{1}{\sqrt{2}}(-1+\frac{Y_{\cal S'}<H_{1_2}>}{Y_{\cal S}})\\
\frac{1}{\sqrt{2}}(-1-\frac{Y_{\cal S'}<H_{1_2}>}{Y_{\cal S}})&2
\end{pmatrix}.
\end{equation}

We can similary give a VEV to $H_{1_3'}$, which will give an up-type matrix:
\begin{equation}
U'=Y_{\cal C}
\begin{pmatrix}
\frac{m_{\mu}}{Y_{\mu}}&-\frac{1}{\sqrt{2}}(\frac{m_{e}}{Y_{e}}-\frac{Y_{\cal C'}<H_{1_3'}>}{Y_{\cal C}})\\
-\frac{1}{\sqrt{2}}(\frac{m_{e}}{Y_{e}}+\frac{Y_{\cal C'}<H_{1_3'}>}{Y_{\cal C}})& \frac{m_{\tau}}{Y_{\tau}}
\end{pmatrix}.
\end{equation}
But because the diagonal elements of $U'$ are so large, this will not have a significant effect. Similar to what the neutrino sector above, we can vary the value $\frac{Y_{\cal S'}<H_{1_2}>}{Y_{\cal S}}$ in order to get a more accurate prediction for the Cabibbo angle. Because we are only varying one parameter we can find the optimal value to a much higher precision. In fact setting the value to $\frac{Y_{\cal S'}<H_{1_2}>}{Y_{\cal S}} = .8792$ gives us a mixing matrix:
\begin{equation}
W=
\begin{pmatrix}
0.974464&-0.224545\\
0.224545&0.974464
\end{pmatrix}.
\end{equation}

The values for $W_{ud}$ and $W_{us}$, and thus the prediction for the Cabibbo angle, are almost identical to those found from the latest experimental fit  \cite{Tanabashi:2018}:

\begin{equation}
|W|=
\begin{pmatrix}
0.97446 \pm 0.00010 & 0.22452 \pm .00044\\
0.22438 \pm 0.00044 & 0.97359^{+0.00010}_{-0.00011}
\end{pmatrix}.
\end{equation}
Specifically, our errors are (again in units of $\sigma$):
\begin{equation}
\begin{pmatrix}
0.0379&0.0562\\
0.3744&7.9435
\end{pmatrix}.
\end{equation}

The $W_{cd}$ prediction is also well within 1$\sigma$. The prediction for $W_{cs}$ is quite a bit off, but this is to be expected, or at least not surprising,  given our neglect of third family mixing effects.

\section{$T'$ Embedding in $SU(2)$}

As explained in the Introduction, it is often desirable to embed discrete symmetries into continuous gauge   groups at higher energy scales. The remainder of this paper will focus on generalizing our $T'$ model to a gauged $SU(2)_{T'}$ flavor theory.

There are three main tasks needed for our gauge group embedding. First, we must identify which $SU(2)_{T'}$ representations our $T'$ particles can fall into. This is easily accomplished by examining the branching rules from Table \ref{tab:2}. New particles will have to be introduced to fill out these $SU(2)_{T'}$ irreps, as a full theory cannot contain incomplete group representations.
Second, we must ensure our theory is anomaly free. This involves checking that our representations satisfy certain sum rules on their quantum numbers (see e.g.,  \cite{Bilal:2008qx}). Again we will see we must add more particles to the theory in order to cancel all anomalies. Finally, we formulate a scalar Lagrangian where we can find a particular vacuum expectation value that breaks $SU(2)_{T'}$ down stepwise to $T'$ \cite{Luhn:2011ip,Merle:2011vy,Rachlin:2017rvm}, then to $Q$, etc. and eventually to nothing.

\subsection{$SU(2)$ Multiplets}

Table I shows the results of embedding the $T'$ irreps of equations (\ref{leptonreps}) and (\ref{quarkreps}) into $\text{SU(2)}_{T'}$. Each row shows the particle in their  $\text{SU(2)}_{T'}$ multiplet, and each column gives the representation of the constituent particles under the specified gauge group.

\begin{table}[H]\label{allreps}
\centering
\caption{Fermionic content of $\text{SU(2)}_{T'}$ gauge theory}
\begin{tabular}{|c||c|c|c|c|}
\hline
Particles & SU(3)& SU(2)& U(1) Charge &$\text{SU(2)}_{T'}$\\
\hline
$((\nu_\tau,\tau),(\nu_\mu,\mu),(\nu_e,e))_L$&1&2&-1&3\\
$\tau_R^c$&1&1&2&1\\
$(A,B,C)_L$&1&1&-2&3\\
$(\mu,e,A,B,C)_R^c$&1&1&2&5\\
$(N_1,N_2,N_3)_R$&1&1&0&3\\
$((c,s),(u,d))_L$&3&2&$\frac{1}{3}$&2\\
$(t,b)_L$&3&2&$\frac{1}{3}$&1\\
$t_R^c$&3&1&$-\frac{4}{3}$&1\\
$(X,b,\alpha,\beta,\gamma)_R^c$&3&1&$\frac{2}{3}$&5\\
$X_L$&3&1&$-\frac{2}{3}$&1\\
$(\alpha,\beta,\gamma)_L$&3&1&$-\frac{2}{3}$&3\\
$(c,u,i,j)_R^c$&3&1&$-\frac{4}{3}$&4\\
$(s,d,k,l)_R^c$&3&1&$\frac{2}{3}$&4\\
$i_L$&3&1&$\frac{4}{3}$&1\\
$j_L$&3&1&$\frac{4}{3}$&1\\
$k_L$&3&1&$-\frac{2}{3}$&1\\
$l_L$&3&1&$-\frac{2}{3}$&1\\
%$H_5$&1&2&0&5\\
%$H_5'$&1&2&0&5\\
%$H_7$&1&1?&0&7\\
\hline

\end{tabular}

\end{table}

\

In order to complete the various irreps of $ SU(2)_{T'}$ we have to include a number of new particles. Specifically we have added three new leptons: $(a,b,c)$, and eight new quarks: $X,\alpha,\beta,\gamma,i,j,k,l$. %All of our scalar particles have slotted neatly into $H_5$ and $H_5'$ and we have added $H_7$ to facilitate the breaking from $T'$ to $SU(2)_{T'}$ (see section [?] below).c

Our next step is to check our theory for anomalies. With the current irreps, the only anomaly that does not cancel is $SU(2)_{T'}\times SU(2)_{T'}\times  U(1)_Y$. To cancel this anomaly and avoid disrupting other cancellations, we add the  multiplets listed in Table II to the theory.   Note that this is not the only way to do the embedding, but it is the most straightforward and economical embedding we have found. 

\begin{table}[H]\label{newparticles}
\centering
\caption{Additional particles needed for cancellation of chiral anomalies} 

\begin{tabular}{|c||c|c|c|c|c|}
\hline
Particles & $SU(3)_C$ irrep& $SU(2)_L$ irrep& $U(1)_Y$ charge &$SU(2)_{T'}$ irrep\\
\hline
$(a_1,a_2,a_3,a_4,a_5)$&1&1&-2&5\\
$(n_1,n_2)$&1&1&2&2\\
$m_1$&1&1&2&1\\
$m_{1'}$&1&1&2&1\\
$m_{1''}$&1&1&2&1\\
\hline
\end{tabular}
\end{table}

With that we have a complete fermion sector for the theory. Although we have had to add many new particles, all of them can be made sufficiently heavy such that they are only relevant at very high energy scales. 
%{\bf Lepton Lagragian $SU(2)$ and $T'$:}
%\begin{equation}
%{\cal L}_{l}= L_L \tau_R H_3 +L_L L_R (H_3+H_5+H_7) + L_L N_R (H_1 +H_3+H_5) +L'_L L_R (H_3+H_5+H_7)

%\end{equation}
%\\
%\begin{equation}
%{\cal L}_{l}= L_L \tau_R H_{3'} +L_L \mu_R H_{3'}+ L_L e_R H_{3'}+ L_L N_R (H_{1_1}+H_{1_2}+H_{1_3}+H_3)+ N_R N_R
%\end{equation}

\subsection{$Z_2$ anomaly cancellation}

In the above formulation we have canceled all anomalies that come about due to the addition of the $T'$ symmetry to the standard model. However, recall that we also included an extra $Z_2$ symmetry in order to forbid certain unwanted Lagrangian terms. This $Z_2$ can be embedded in an extra $U(1)_{Z_2}$ symmetry that breaks at an arbitrary scale independent of the $SU(2)_{T'}$ breaking. We detail the charge assignments for an example anomaly-free $SU(3)\times SU(2)_L \times U(1)_Y \times SU(2)_{T'} \times U(1)_{Z_2}$ theory below in Table III. Notice we have added an SM singlet $SU(2)_{T'}$ {\bf 4} with $Z_2$  charge $-1$ and fourteen fermions that are trivial singlets under everything but $U(1)_Y\times U(1)_{Z_2}$. Five of them, the $E$s have charge (2,1) and the other five, the $F$s have charge ($-2$,0) under this group, the remaining four have $U(1)_Y$ charge $\pm 10$ and $U(1)_{Z_2}$ charge 0, 1 or -1.

There is significant freedom in assigning $U(1)_{Z_2}$ charges to existing particles as  they reduce to particles with identical $Z_2$ charges modulo 2. So even though this example has involved adding many extra particles, a less baroque model may be possible.

\begin{table}[H]\label{Z2reps}
\label{z2reps}
\centering
\caption{Example charge assignments for $U(1)_{Z_2}$ anomaly cancellation  \cite{Ibanez:1991pr,Ibanez:1991hv}}
\begin{tabular}{|c||c|c|c|c|c|}
\hline
Particles & $SU(3)_C$ irrep& $SU(2)_L$ irrep& $U(1)_Y$ charge &$SU(2)_{T'}$ irrep&$\text{U(1)}_{Z_2}$ charge\\
\hline
$((\nu_\tau,\tau),(\nu_\mu,\mu),(\nu_e,e))_L$&1&2&-1&3&0\\
$\tau_R^c$&1&1&2&1&1\\
$(A,B,C)_L$&1&1&-2&3&0\\
$(\mu,e,A,B,C)_R^c$&1&1&2&5&-1\\
$(N_1,N_2,N_3)_R$&1&1&0&3&0\\
$((c,s),(u,d))_L$&3&2&$\frac{1}{3}$&2&0\\
$(t,b)_L$&3&2&$\frac{1}{3}$&1&0\\
$t_R^c$&3&1&$-\frac{4}{3}$&1&0\\
$(X,b,\alpha,\beta,\gamma)_R^c$&3&1&$\frac{2}{3}$&5&1\\
$X_L$&3&1&$-\frac{2}{3}$&1&-1\\
$(\alpha,\beta,\gamma)_L$&3&1&$-\frac{2}{3}$&3&0\\
$(c,u,i,j)_R^c$&3&1&$-\frac{4}{3}$&4&-1\\
$(s,d,k,l)_R^c$&3&1&$\frac{2}{3}$&4&0\\
$i_L$&3&1&$\frac{4}{3}$&1&0\\
$j_L$&3&1&$\frac{4}{3}$&1&0\\
$k_L$&3&1&$-\frac{2}{3}$&1&0\\
$l_L$&3&1&$-\frac{2}{3}$&1&0\\
$(a_1,a_2,a_3,a_4,a_5)$&1&1&-2&5&0\\
$(n_1,n_2)$&1&1&2&2&0\\
$m_1$&1&1&2&1&1\\
$m_{1'}$&1&1&2&1&1\\
$m_{1''}$&1&1&2&1&1\\
$(b_1,b_2,b_3,b_4)$&1&1&0&4&-1\\
$5\times E$&1&1&2&1&1\\
$5\times F$&1&1&-2&1&0\\
$g_1$&1&1&-10&1&1\\
$g_2$&1&1&-10&1&-1\\
$h_1$&1&1&10&1&0\\
$h_2$&1&1&10&1&0\\
\hline
\end{tabular}
\end{table}

\section{Spontaneous Symmetry Breaking}

Our final step is to provide the spontaneous breaking of $\text{SU(2)}_{T'} \rightarrow T'$. We have already performed this analysis in a previous paper \cite{Rachlin:2017rvm}, so will only summarize the results here. To have this spontaneous symmetry breaking we must include a scalar multiplet of $SU(2)_{T'}$ that contains a trivial singlet of $T'$. Looking at the branching rules of table \ref{tab:2}, we see the smallest avalable irrep for this purpose is the \textbf{7}. The \textbf{7} can be real or complex, but for simplicity we choose a real multiplet with scalar potential  
\begin{equation}\label{scalarlag}
V_7 = -m^2 \,T_{abc}T_{abc}+\lambda \,(T_{abc}T_{abc})^2 +\kappa \,\,T_{abd}T_{abe}T_{fge}T_{fgd},
\end{equation}
where $T$ is a traceless, symmetric, $3\times 3\times 3$ tensor, $\lambda$ and $\kappa$ are the scalar quartic coupling constants, and the indices $a,b,...$ run from 1 to 3.

Spontaneous breaking to $T'$ occurs when the potential is minimized and the scalar is given a Vacuum Expectation Value (VEV) in a particular direction. For the real \textbf{7} this VEV is \cite{Rachlin:2017rvm}:
\begin{equation} \label{eq:so3a4vev}
{\cal V}=\sqrt{\frac{3m^2}{2(3\lambda+\kappa)}}\,[0,0,0,0,0,0,1].
\end{equation}

After breaking  $\text{SU(2)}_{T'}$ the  {\bf 7} real scalars reduce to their $T'$ irreps with mass eigenvalues given by
\begin{table}[H]
\begin{center}
\begin{tabular}{c|c} 
Value & Multiplicity \\ [1ex] 
\hline\hline
0 & 3\\[2ex]
$4m^2$ & 1\\[2ex]
$\frac{8m^2\kappa}{5(3\lambda+\kappa)}$ & 3\\[2ex]
\end{tabular}
\label{tab:A4/7}
\end{center}
\end{table}
\noindent which contains the three requisite Goldstone bosons that get eaten by the $SU(2)_{T'}$ gauge bosons.
To ensure a stable minimum the coupling constants must satisfy the constraints
$3\lambda+\kappa>0$ 
and 
$\kappa > 0$. Clearly there is a substantial region of parameter space where this pattern of SSB is the stable minimum of the potential in eq.(\ref{scalarlag}).

\indent This \textbf{7} is obviously not the only scalar in the theory as more scalars are needed to construct Yukawa terms at the $\text{SU(2)}_{T'}$ scale. 
However, we omit the full scalar Lagrangian in this paper because we will not be exploring  its complete phenomenology at present.
We are assuming that the coupling of the  \textbf{7} to the other scalars is sufficiently weak that the breaking to $T'$ is not destabilized. The analysis of a specific example of this type of VEV stability can be found in \cite{Rachlin:2017rvm}.
\section{Conclusion}

We have extended the basic $T'$ flavor model to fit the current best available quark and lepton mass and mixing angle data. 
More specifically, we have constructed an extended but fairly simple, renormalizable $T'$ model that predicts neutrino mixing parameters within 2$\sigma$ of experiment, as well as a Cabibbo angle well within 1$\sigma$. This has required the addition of $T'$ scalar singlets with VEVs. Once our new $T'$ model was fixed, we then extended it further by embedding it in $SU(2)_{T'}$ such that the entire model was fully gauged. This avoided all problems with gauge and gravity mixed anomalies at the expense of adding a number of new fermions to the lepton and quark sectors. The additional fermions were not necessarily the minimal set, as there are many possible choices, so what we have provided is a proof of principle that fully gauged flavor models can be found to fit all current flavored data. It still remains quite challenging to find a full gauge unification of flavor, but it is perhaps not unreasonable to hope that one could eventually find a top-down GUT flavor model that reduces to a product gauge model of the type we have discussed here.

Besides the $T'$ model discussed here, gauged $A_4$ models \cite{Berger:2009tt,Grossman:2014oqa,King:2018fke} have also appeared, but
there remains a long list of discrete groups   $S_4, A_5,Q_6,O',I',T_7,\Delta(27)$ and $PSL(2,7)$ that are easy to obtain from breaking $SU(2)$
or $SU(3)$. So it appears possible to gauge some if not all of the models based on these groups \cite{Everett:2010rd,Chen:2011dn,Luhn:2007sy,Kile:2014kya,Vien:2015koa,Vien:2016qbb,Vien:2016tmh,Ferreira:2012ri,Chen:2014wiw}.

There is still more to explore within our present model; in particular the phenomenology of the new scalar singlets and the additional fermions required for anomaly cancellation. The phenomenology of the $SU(2)_{T'}$ scalar  \textbf{7} would also benefit from further study, but we leave these topics for future work. Beyond this specific model, it would be preferable to avoid $Z_n$ factors  by either reassigning irreps of SM states, or by using different initial nonabelian discrete groups. This would simplify the anomaly cancelation and hence minimize the introduction of extra fermionic states. We plan to search for such models in the future.

\section{Acknowledgements} 

We thank Jim Talbert for a helpful discussion about discrete anomalies. The work of TWK was supported by DoE grant \# DE-SC-0019235 and that of BLR by DoE grant \# DE-SC-0011981.

\appendix

\section{Useful Information About the Binary Tetrahedral Group $T'$}
\subsection{$T'$ Character Table}
\begin{table}[H]
\centering
\begin{tabular}{c|ccccccc}
Dimension & $\mathbf{C_1}$&$\mathbf{C_2}$&$4\mathbf{C_3}$&$6\mathbf{C_4}$&$4\mathbf{C_5}$&$4\mathbf{C_6}$&$4\mathbf{C_7}$\\
\hline
$\mathbf{1_1}$&1&1&1&1&1&1&1\\
$\mathbf{1_2}$&1&1&$\omega^2$&$\omega^4$&1&$\omega^2$&$\omega^4$\\
$\mathbf{1_3}$&1&1&$\omega^4$&$\omega^2$&1&$\omega^4$&$\omega^2$\\
$\mathbf{2_1}$&2&-2&-1&-1&0&1&1\\
$\mathbf{2_2}$&2&-2&$\omega^5$&$\omega$&0&$\omega^2$&$\omega^4$\\
$\mathbf{2_3}$&2&-2&$\omega$&$\omega^5$&0&$\omega^4$&$\omega^2$\\
$\mathbf{3}$&3&3&0&0&-1&0&0
\end{tabular}
\end{table}

Where $\omega=e^{\frac{2\pi i}{6}}$.
\subsection{Kronecker Products of $T'$ Irreps}
\begin{table}[H]
\centering
\scalebox{.9}{
\begin{tabular}{c||c|c|c|c|c|c|c}
Dimension & $\mathbf{1_1}$&$\mathbf{1_2}$&$\mathbf{1_3}$&$\mathbf{2_1}$&$\mathbf{2_2}$&$\mathbf{2_3}$&$\mathbf{3}$\\
\hline
\hline
$\mathbf{1_1}$&$\mathbf{1_1}$&$\mathbf{1_2}$&$\mathbf{1_3}$&$\mathbf{2_1}$&$\mathbf{2_2}$&$\mathbf{2_3}$&$\mathbf{3}$\\
\hline
$\mathbf{1_2}$&$\mathbf{1_2}$&$\mathbf{1_3}$&$\mathbf{1_1}$&$\mathbf{2_2}$&$\mathbf{2_3}$&$\mathbf{2_1}$&$\mathbf{3}$\\
\hline
$\mathbf{1_3}$&$\mathbf{1_3}$&$\mathbf{1_1}$&$\mathbf{1_2}$&$\mathbf{2_3}$&$\mathbf{2_1}$&$\mathbf{2_2}$&$\mathbf{3}$\\
\hline
$\mathbf{2_1}$&$\mathbf{2_1}$&$\mathbf{2_2}$&$\mathbf{2_3}$&$\mathbf{1_1}+\mathbf{3}$&$\mathbf{1_2}+\mathbf{3}$&$\mathbf{1_3}+\mathbf{3}$&$\mathbf{2_1}+\mathbf{2_2}+\mathbf{2_3}$\\
\hline
$\mathbf{2_2}$&$\mathbf{2_2}$&$\mathbf{2_3}$&$\mathbf{2_1}$&$\mathbf{1_2}+\mathbf{3}$&$\mathbf{1_3}+\mathbf{3}$&$\mathbf{1_1}+\mathbf{3}$&$\mathbf{2_1}+\mathbf{2_2}+\mathbf{2_3}$\\
\hline
$\mathbf{2_3}$&$\mathbf{2_3}$&$\mathbf{2_1}$&$\mathbf{2_2}$&$\mathbf{1_3}+\mathbf{3}$&$\mathbf{1_1}+\mathbf{3}$&$\mathbf{1_2}+\mathbf{3}$&$\mathbf{2_1}+\mathbf{2_2}+\mathbf{2_3}$\\
\hline
$\mathbf{3}$&$\mathbf{3}$&$\mathbf{3}$&$\mathbf{3}$&$\mathbf{2_1}+\mathbf{2_2}+\mathbf{2_3}$&$\mathbf{2_1}+\mathbf{2_2}+\mathbf{2_3}$&$\mathbf{2_1}+\mathbf{2_2}+\mathbf{2_3}$&$\mathbf{1_1}+\mathbf{1_2}+\mathbf{1_3}+\mathbf{3}+\mathbf{3}$
\end{tabular}
}
\end{table}
\subsection{Decomposition of SU(2) Irreps to $T'$ Irreps}

\begin{table}[H]
\caption{$SU(2)\rightarrow T'$}
\label{tab:2}
\begin{center}
\begin{tabular}{c|c|ccccccc}
SU(2) &Dynkin Index&&&&$T'$\\
\hline
& & $\mathbf{1_1}$&$\mathbf{1_2}$&$\mathbf{1_3}$&$\mathbf{2_1}$&$\mathbf{2_2}$&$\mathbf{2_3}$&$\mathbf{3}$\\
\hline
\textbf{1}&0&1&0&0&0&0&0&0\\
\textbf{2}&1&0&0&0&1&0&0&0\\
\textbf{3}&4&0&0&0&0&0&0&1\\
\textbf{4}&10&0&0&0&0&1&1&0\\
\textbf{5}&20&0&1&1&0&0&0&1\\
\textbf{6}&35&0&0&0&1&1&1&0\\
\textbf{7}&56&1&0&0&0&0&0&2\\
\textbf{8}&84&0&0&0&2&1&1&0\\
\textbf{9}&120&1&1&1&0&0&0&2
\end{tabular}
  \end{center}
\end{table}

\subsection{$T'$ Clebsch-Gordan Coefficients}
For our basis we take the tensor products in section 5 of \cite{Ishimori:2010au} with $p=i$, $p_1=-1$, and $p_2=1$.

\begin{eqnarray}
\2tvec{x_1}{x_2}_{\bf 2(2')}\otimes\2tvec{y_1}{y_2}_{\bf 2(2'')}
=
\left(\frac{x_1y_2-x_2y_1}{\sqrt{2}}\right)_{\bf 1}
\oplus\3tvec{\frac{-1}{\sqrt{2}}(x_1y_2+x_2y_1)}{-x_1y_1}{x_2y_2}_{\bf3},
\end{eqnarray}
\begin{eqnarray}
\2tvec{x_1}{x_2}_{\bf2'(2)}\otimes\2tvec{y_1}{y_2}_{\bf2'(2'')}
&=\left(\frac{x_1y_2-x_2y_1}{\sqrt{2}}\right)_{\bf1''}
\oplus
\3tvec{x_1y_1}{x_2y_2}{\frac{1}{\sqrt{2}}(x_1y_2+x_2y_1)}_{\bf3}, 
\\
\2tvec{x_1}{x_2}_{\bf2''(2)}\otimes\2tvec{y_1}{y_2}_{\bf2''(2')}
&=\left(\frac{x_1y_2-x_2y_1}{\sqrt{2}}\right)_{\bf1'}
\oplus
\3tvec{x_2y_2}{\frac{-1}{\sqrt{2}}(x_1y_2+x_2y_1)}{x_1y_1}_{\bf3},
\end{eqnarray}

\begin{eqnarray}
\3tvec{x_1}{x_2}{x_3}_{\bf3}\otimes\3tvec{y_1}{y_2}{y_3}_{\bf3}
&=&
[x_1y_1+x_2y_3+x_3y_2]_{\bf1}\nonumber
\\&\oplus&
[x_3y_3 -(x_1y_2+x_2y_1)]_{\bf1'}
\oplus
[(x_2y_2-(x_1y_3+x_3y_1)]_{\bf1''}\nonumber
\\&\oplus&
\3tvec{2x_1y_1-x_2y_3-x_3y_2)}
{-2x_3y_3-x_1y_2-x_2y_1}{-2x_2y_2-x_1y_3-x_3y_1}_{\bf3}
\nonumber
\\&\oplus&
\3tvec{x_2y_3-x_3y_2}{x_1y_2-x_2y_1}
{x_3y_1-x_1y_3}_{\bf3},
\end{eqnarray}
%%%%%%%%%%%%%%%%%%%%%%%%%%%%%%
\begin{eqnarray}
\2tvec{x_1}{x_2}_{\bf2,2',2''}\otimes\3tvec{y_1}{y_2}{y_3}_{\bf3}
&=&
  \left(\begin{array}{c}
 -\sqrt2x_2y_2+x_1y_1
\\ 
-\sqrt2x_1y_3-x_2y_1
\end{array}\right)_{\bf2,2',2''}
\nonumber\\&\oplus &
 \left(\begin{array}{c}
\sqrt2x_2y_3+x_1y_2
\\
 -\sqrt2x_1y_1-x_2y_2
\end{array}\right)_{\bf2',2'',2}
\nonumber\\&\oplus&
 \left(\begin{array}{c}
 -\sqrt2x_2y_1+x_1y_3
 \\
\sqrt2x_1y_2-x_2y_3
\end{array}\right)_{\bf2'',2,2'},
%%%%%
\end{eqnarray}
%%%%%%%%%%%%%%%
\begin{eqnarray}
&&(x)_{\bf1'(1'')}\otimes\2tvec{y_1}{y_2}_{\bf2,2',2''}
=
\2tvec{xy_1}{xy_2}_{\bf2'(2''),2''(2),2(2')},\\
&&(x)_{\bf1'}\otimes\3tvec{y_1}{y_2}{y_3}_{\bf3}
=
\3tvec{xy_3}{xy_1}{-xy_2}_{\bf3},\quad
(x)_{\bf1''}\otimes\3tvec{y_1}{y_2}{y_3}_{\bf3}
=
\3tvec{xy_2}{-xy_3}{xy_1}_{\bf3}.
\end{eqnarray}
%%%


\newpage

 



\begin{thebibliography}{99}

  %HISTORY


  


  %REVIEWS
%\cite{Frampton:1994rk,Ishimori:2010au,Altarelli:2010gt,King:2013eh,King:2014nza}
  
\bibitem{Frampton:1994rk}
  P.~H.~Frampton and T.~W.~Kephart,
  %``Simple nonAbelian finite flavor groups and fermion masses,''
  Int.\ J.\ Mod.\ Phys.\  A {\bf 10}, 4689 (1995)
  [arXiv:hep-ph/9409330].
  
\bibitem{Ishimori:2010au}
  H.~Ishimori, T.~Kobayashi, H.~Ohki, H.~Okada, Y.~Shimizu and M.~Tanimoto,
  %``Non-Abelian Discrete Symmetries in Particle Physics,''
  Prog.\ Theor.\ Phys.\ Suppl.\  {\bf 183}, 1 (2010)
  [arXiv:1003.3552 [hep-th]].  
  
  %\cite{Altarelli:2010gt}
\bibitem{Altarelli:2010gt} 
  G.~Altarelli and F.~Feruglio,
  %``Discrete Flavor Symmetries and Models of Neutrino Mixing,''
  Rev.\ Mod.\ Phys.\  {\bf 82}, 2701 (2010)
  doi:10.1103/RevModPhys.82.2701
  [arXiv:1002.0211 [hep-ph]].
  %%CITATION = doi:10.1103/RevModPhys.82.2701;%%
  %495 citations counted in INSPIRE as of 09 Oct 2016
  
    
 %\cite{King:2013eh}
\bibitem{King:2013eh} 
  S.~F.~King and C.~Luhn,
  %``Neutrino Mass and Mixing with Discrete Symmetry,''
  Rept.\ Prog.\ Phys.\  {\bf 76}, 056201 (2013)
  doi:10.1088/0034-4885/76/5/056201
  [arXiv:1301.1340 [hep-ph]].
  %%CITATION = doi:10.1088/0034-4885/76/5/056201;%%
  %331 citations counted in INSPIRE as of 08 Feb 2017
  
  %\cite{King:2014nza}
\bibitem{King:2014nza} 
  S.~F.~King, A.~Merle, S.~Morisi, Y.~Shimizu and M.~Tanimoto,
  %``Neutrino Mass and Mixing: from Theory to Experiment,''
  New J.\ Phys.\  {\bf 16}, 045018 (2014)
  doi:10.1088/1367-2630/16/4/045018
  [arXiv:1402.4271 [hep-ph]].
  %%CITATION = doi:10.1088/1367-2630/16/4/045018;%%
  %150 citations counted in INSPIRE as of 23 Feb 2017
  
  % top down
  
  %\cite{Albright:2016lpi}
\bibitem{Albright:2016lpi} 
  C.~H.~Albright, R.~P.~Feger and T.~W.~Kephart,
  %``Unification of Gauge, Family, and Flavor Symmetries Illustrated in Gauged SU(12) Models,''
  Phys.\ Rev.\ D {\bf 93}, no. 7, 075032 (2016)
  doi:10.1103/PhysRevD.93.075032
  [arXiv:1601.07523 [hep-ph]].
  %%CITATION = doi:10.1103/PhysRevD.93.075032;%%
  %2 citations counted in INSPIRE as of 15 Nov 2018
  

  
   
  %T '
%\cite{Frampton:1994rk,Aranda:1999kc,Chen:2007afa,Frampton:2007et,Frampton:2008bz,Frampton:2010uw,Natale:2016xob,Carone:2016xsi}
  
  %\cite{Aranda:1999kc}
\bibitem{Aranda:1999kc} 
  A.~Aranda, C.~D.~Carone and R.~F.~Lebed,
  %``U(2) flavor physics without U(2) symmetry,''
  Phys.\ Lett.\ B {\bf 474}, 170 (2000)
  doi:10.1016/S0370-2693(99)01497-5
  [hep-ph/9910392].
  %%CITATION = doi:10.1016/S0370-2693(99)01497-5;%%
  %80 citations counted in INSPIRE as of 16 Feb 2017
  
   %\cite{Chen:2007afa}
\bibitem{Chen:2007afa} 
  M.~C.~Chen and K.~T.~Mahanthappa,
  %``CKM and Tri-bimaximal MNS Matrices in a $SU(5) \times ^{(d)}T$ Model,''
  Phys.\ Lett.\ B {\bf 652}, 34 (2007)
  doi:10.1016/j.physletb.2007.06.064
  [arXiv:0705.0714 [hep-ph]].
  %%CITATION = doi:10.1016/j.physletb.2007.06.064;%%
  %207 citations counted in INSPIRE as of 19 Feb 2017
  
  \bibitem{Frampton:2007et}
  P.~H.~Frampton and T.~W.~Kephart, %``Flavor Symmetry for Quarks and Leptons,''
  JHEP {\bf 0709}, 110 (2007)
  [arXiv:0706.1186 [hep-ph]];
 % \bibitem{Frampton:2008bz}
 

 
%\cite{Frampton:2008bz}
\bibitem{Frampton:2008bz} 
  P.~H.~Frampton, T.~W.~Kephart and S.~Matsuzaki,
  %``Simplified Renormalizable T-prime Model for Tribimaximal Mixing and Cabibbo Angle,''
  Phys.\ Rev.\ D {\bf 78}, 073004 (2008)
  doi:10.1103/PhysRevD.78.073004
  [arXiv:0807.4713 [hep-ph]].
  %%CITATION = doi:10.1103/PhysRevD.78.073004;%%
  %55 citations counted in INSPIRE as of 18 Feb 2017
  
  %\cite{Frampton:2010uw}
\bibitem{Frampton:2010uw} 
  P.~H.~Frampton, C.~M.~Ho, T.~W.~Kephart and S.~Matsuzaki,
  %``LHC Higgs Production and Decay in the $T'$ Model,''
  Phys.\ Rev.\ D {\bf 82}, 113007 (2010)
  doi:10.1103/PhysRevD.82.113007
  [arXiv:1009.0307 [hep-ph]].
  %%CITATION = doi:10.1103/PhysRevD.82.113007;%%
  %14 citations counted in INSPIRE as of 18 Feb 2017

  %\cite{Natale:2016xob}
\bibitem{Natale:2016xob} 
  A.~Natale,
  %``A Radiative Model of Quark Masses with Binary Tetrahedral Symmetry,''
  Nucl.\ Phys.\ B {\bf 914}, 201 (2017)
  doi:10.1016/j.nuclphysb.2016.11.006
  [arXiv:1608.06999 [hep-ph]].
  %%CITATION = doi:10.1016/j.nuclphysb.2016.11.006;%%
  
   %\cite{Carone:2016xsi}
\bibitem{Carone:2016xsi} 
  C.~D.~Carone, S.~Chaurasia and S.~Vasquez,
  %``Flavor from the double tetrahedral group without supersymmetry,''
  Phys.\ Rev.\ D {\bf 95}, no. 1, 015025 (2017)
  doi:10.1103/PhysRevD.95.015025
  [arXiv:1611.00784 [hep-ph]].
  %%CITATION = doi:10.1103/PhysRevD.95.015025;%%
  
  %data fit
  
  %\cite{Esteban:2016qun}
\bibitem{Esteban:2016qun} 
  I.~Esteban, M.~C.~Gonzalez-Garcia, M.~Maltoni, I.~Martinez-Soler and T.~Schwetz,
  %``Updated fit to three neutrino mixing: exploring the accelerator-reactor complementarity,''
  JHEP {\bf 1701}, 087 (2017)
  doi:10.1007/JHEP01(2017)087
  [arXiv:1611.01514 [hep-ph]].
  %%CITATION = doi:10.1007/JHEP01(2017)087;%%
  %398 citations counted in INSPIRE as of 15 Nov 2018
  
  % grav and discrete sym
  
    %\cite{Krauss:1988zc}
\bibitem{Krauss:1988zc} 
  L.~M.~Krauss and F.~Wilczek,
  %``Discrete Gauge Symmetry in Continuum Theories,''
  Phys.\ Rev.\ Lett.\  {\bf 62}, 1221 (1989).
  doi:10.1103/PhysRevLett.62.1221
  %%CITATION = doi:10.1103/PhysRevLett.62.1221;%%
  %488 citations counted in INSPIRE as of 16 Feb 2017
  
  %Discrete anomalies
  
 % \cite{Ibanez:1991pr,Ibanez:1991hv,Luhn:2008sa,Fallbacher:2015pga,Talbert:2018nkq}
  
  %\cite{Ibanez:1991pr}
\bibitem{Ibanez:1991pr} 
  L.~E.~Ibanez and G.~G.~Ross,
  %``Discrete gauge symmetries and the origin of baryon and lepton number conservation in supersymmetric versions of the standard model,''
  Nucl.\ Phys.\ B {\bf 368}, 3 (1992).
  doi:10.1016/0550-3213(92)90195-H
  %%CITATION = doi:10.1016/0550-3213(92)90195-H;%%
  %573 citations counted in INSPIRE as of 15 Nov 2018
  
  %\cite{Ibanez:1991hv}
\bibitem{Ibanez:1991hv} 
  L.~E.~Ibanez and G.~G.~Ross,
  %``Discrete gauge symmetry anomalies,''
  Phys.\ Lett.\ B {\bf 260}, 291 (1991).
  doi:10.1016/0370-2693(91)91614-2
  %%CITATION = doi:10.1016/0370-2693(91)91614-2;%%
  %351 citations counted in INSPIRE as of 15 Nov 2018
  
   %\cite{Luhn:2008sa}
\bibitem{Luhn:2008sa} 
  C.~Luhn and P.~Ramond,
  %``Anomaly Conditions for Non-Abelian Finite Family Symmetries,''
  JHEP {\bf 0807}, 085 (2008)
  doi:10.1088/1126-6708/2008/07/085
  [arXiv:0805.1736 [hep-ph]].
  %%CITATION = doi:10.1088/1126-6708/2008/07/085;%%
  %32 citations counted in INSPIRE as of 16 Feb 2017
  
  %\cite{Fallbacher:2015pga}
\bibitem{Fallbacher:2015pga} 
  M.~Fallbacher,
  %``Breaking classical Lie groups to finite subgroups Ð an automated approach,''
  Nucl.\ Phys.\ B {\bf 898}, 229 (2015)
  doi:10.1016/j.nuclphysb.2015.07.004
  [arXiv:1506.03677 [hep-th]].
  %%CITATION = doi:10.1016/j.nuclphysb.2015.07.004;%%
  %2 citations counted in INSPIRE as of 06 Oct 2016
  
  %\cite{Talbert:2018nkq}
\bibitem{Talbert:2018nkq} 
  J.~Talbert,
  %``Pocket Formulae for Non-Abelian Discrete Anomaly Freedom,''
  Phys.\ Lett.\ B {\bf 786}, 426 (2018)
  doi:10.1016/j.physletb.2018.10.025
  [arXiv:1804.04237 [hep-ph]].
  %%CITATION = doi:10.1016/j.physletb.2018.10.025;%%
  %1 citations counted in INSPIRE as of 15 Nov 2018
  
  %A4

%\cite{Ma:2001dn,Babu:2002dz}
  
  %\cite{Ma:2001dn}
\bibitem{Ma:2001dn} 
  E.~Ma and G.~Rajasekaran,
  %``Softly broken A(4) symmetry for nearly degenerate neutrino masses,''
  Phys.\ Rev.\ D {\bf 64}, 113012 (2001)
  doi:10.1103/PhysRevD.64.113012
  [hep-ph/0106291].
  %%CITATION = doi:10.1103/PhysRevD.64.113012;%%
  %605 citations counted in INSPIRE as of 16 Feb 2017
  
  %\cite{Babu:2002dz}
\bibitem{Babu:2002dz} 
  K.~S.~Babu, E.~Ma and J.~W.~F.~Valle,
  %``Underlying A(4) symmetry for the neutrino mass matrix and the quark mixing matrix,''
  Phys.\ Lett.\ B {\bf 552}, 207 (2003)
  doi:10.1016/S0370-2693(02)03153-2
  [hep-ph/0206292].
  %%CITATION = doi:10.1016/S0370-2693(02)03153-2;%%
  %590 citations counted in INSPIRE as of 16 Feb 2017
  
  %\cite{PMNS}
  \bibitem{PMNS}  
  B.  Pontecorvo, Soviet Physics JETP. 7: 172. 1958; Z. Maki,  M. Nakagawa, S. Sakata,   
  %"Remarks on the Unified Model of Elementary Particles"
  Progress of Theoretical Physics. 28 (5): 870 (1962).
  
  %\cite{Harrison:2002er}
\bibitem{Harrison:2002er} 
  P.~F.~Harrison, D.~H.~Perkins and W.~G.~Scott,
  %``Tri-bimaximal mixing and the neutrino oscillation data,''
  Phys.\ Lett.\ B {\bf 530}, 167 (2002)
  doi:10.1016/S0370-2693(02)01336-9
  [hep-ph/0202074].
  %%CITATION = doi:10.1016/S0370-2693(02)01336-9;%%
  %1378 citations counted in INSPIRE as of 16 Nov 2018
  
  %\cite{Rahat:2018sgs}
\bibitem{Rahat:2018sgs} 
  M.~H.~Rahat, P.~Ramond and B.~Xu,
  %``Asymmetric tribimaximal texture,''
  Phys.\ Rev.\ D {\bf 98}, no. 5, 055030 (2018)
  doi:10.1103/PhysRevD.98.055030
  [arXiv:1805.10684 [hep-ph]].
  %%CITATION = doi:10.1103/PhysRevD.98.055030;%%
  %1 citations counted in INSPIRE as of 14 Dec 2018
  
  %\cite{Eby:2008uc}
\bibitem{Eby:2008uc} 
  D.~A.~Eby, P.~H.~Frampton and S.~Matsuzaki,
  %``Predictions of Neutrino Mixing Angles in a T-prime Model,''
  Phys.\ Lett.\ B {\bf 671}, 386 (2009)
  doi:10.1016/j.physletb.2008.11.074
  [arXiv:0810.4899 [hep-ph]].
  %%CITATION = doi:10.1016/j.physletb.2008.11.074;%%
  %36 citations counted in INSPIRE as of 16 Nov 2018
  
  %\cite{Esteban:2016qun}
\bibitem{Esteban:2016qun} 
  I.~Esteban, M.~C.~Gonzalez-Garcia, M.~Maltoni, I.~Martinez-Soler and T.~Schwetz,
  %``Updated fit to three neutrino mixing: exploring the accelerator-reactor complementarity,''
  JHEP {\bf 1701}, 087 (2017)
  doi:10.1007/JHEP01(2017)087
  [arXiv:1611.01514 [hep-ph]].
  %%CITATION = doi:10.1007/JHEP01(2017)087;%%
  %402 citations counted in INSPIRE as of 21 Nov 2018
  
  %\cite{NuFIT2018}
\bibitem{NuFIT2018}
  NuFIT 3.2 (2018), www.nu-fit.org.
  
    %\cite{Tanabashi:2018}
  \bibitem{Tanabashi:2018}
  M. Tanabashi et al. (Particle Data Group), Phys. Rev. D 98, 030001 (2018)
  
  %\cite{Bilal:2008qx}
\bibitem{Bilal:2008qx} 
  A.~Bilal,
  %``Lectures on Anomalies,''
  arXiv:0802.0634 [hep-th].
  %%CITATION = ARXIV:0802.0634;%%
  %74 citations counted in INSPIRE as of 04 Sep 2018
  
    %\cite{Luhn:2011ip,Merle:2011vy,Rachlin:2017rvm}
  
  %\cite{Luhn:2011ip}
\bibitem{Luhn:2011ip} 
  C.~Luhn,
  %``Spontaneous breaking of SU(3) to finite family symmetries: a pedestrian's approach,''
  JHEP {\bf 1103}, 108 (2011)
  doi:10.1007/JHEP03(2011)108
  [arXiv:1101.2417 [hep-ph]].
  %%CITATION = doi:10.1007/JHEP03(2011)108;%%
  %25 citations counted in INSPIRE as of 19 Oct 2016
  
  %\cite{Merle:2011vy}
\bibitem{Merle:2011vy} 
  A.~Merle and R.~Zwicky,
  %``Explicit and spontaneous breaking of SU(3) into its finite subgroups,''
  JHEP {\bf 1202}, 128 (2012)
  doi:10.1007/JHEP02(2012)128
  [arXiv:1110.4891 [hep-ph]].
  %%CITATION = doi:10.1007/JHEP02(2012)128;%%
  %27 citations counted in INSPIRE as of 06 Oct 2016

  %\cite{Rachlin:2017rvm}
\bibitem{Rachlin:2017rvm} 
  B.~L.~Rachlin and T.~W.~Kephart,
  %``Spontaneous Breaking of Gauge Groups to Discrete Symmetries,''
  JHEP {\bf 1708}, 110 (2017)
  doi:10.1007/JHEP08(2017)110
  [arXiv:1702.08073 [hep-ph]].
  %%CITATION = doi:10.1007/JHEP08(2017)110;%%
  %1 citations counted in INSPIRE as of 04 Sep 2018
  
  
    %\cite{Berger:2009tt,Grossman:2014oqa,King:2018fke}
  
    %\cite{Berger:2009tt}
\bibitem{Berger:2009tt} 
  J.~Berger and Y.~Grossman,
  %``Model of leptons from SO(3) ---> A(4),''
  JHEP {\bf 1002}, 071 (2010)
  doi:10.1007/JHEP02(2010)071
  [arXiv:0910.4392 [hep-ph]].
  %%CITATION = doi:10.1007/JHEP02(2010)071;%%
  %46 citations counted in INSPIRE as of 13 Sep 2018
  
%\cite{Grossman:2014oqa}
\bibitem{Grossman:2014oqa} 
  Y.~Grossman and W.~H.~Ng,
  %``Nonzero $\theta_{13}$ in $SO(3) \rightarrow A_4$ lepton models,''
  Phys.\ Rev.\ D {\bf 91}, no. 7, 073005 (2015)
  doi:10.1103/PhysRevD.91.073005
  [arXiv:1404.1413 [hep-ph]].
  %%CITATION = doi:10.1103/PhysRevD.91.073005;%%
  %3 citations counted in INSPIRE as of 13 Sep 2018

  
  %\cite{King:2018fke}
\bibitem{King:2018fke} 
  S.~F.~King and Y.~L.~Zhou,
  %``Spontaneous breaking of $SO(3)$ to finite family symmetries with supersymmetry - an $A_4$ model,''
  arXiv:1809.10292 [hep-ph].
  %%CITATION = ARXIV:1809.10292;%%
  %4 citations counted in INSPIRE as of 07 Nov 2018
  
 % O'
  
  %I '
  
   %\cite{Everett:2010rd,Chen:2011dn,Luhn:2007sy,Kile:2014kya,Vien:2015koa,Vien:2016qbb,Vien:2016tmh,Ferreira:2012ri,Chen:2014wiw}

  
  %\cite{Everett:2010rd}
\bibitem{Everett:2010rd} 
  L.~L.~Everett and A.~J.~Stuart,
  %``The Double Cover of the Icosahedral Symmetry Group and Quark Mass Textures,''
  Phys.\ Lett.\ B {\bf 698}, 131 (2011)
  doi:10.1016/j.physletb.2011.02.054
  [arXiv:1011.4928 [hep-ph]].
  %%CITATION = doi:10.1016/j.physletb.2011.02.054;%%
  %20 citations counted in INSPIRE as of 09 Oct 2016
  
  %\cite{Chen:2011dn}
\bibitem{Chen:2011dn} 
  C.~S.~Chen, T.~W.~Kephart and T.~C.~Yuan,
  %``Binary Icosahedral Flavor Symmetry for Four Generations of Quarks and Leptons,''
  PTEP {\bf 2013}, no. 10, 103B01 (2013)
  doi:10.1093/ptep/ptt071
  [arXiv:1110.6233 [hep-ph]].
  %%CITATION = doi:10.1093/ptep/ptt071;%%
  %6 citations counted in INSPIRE as of 09 Oct 2016
  

  
  %T7
  %\cite{Luhn:2007sy}
\bibitem{Luhn:2007sy} 
  C.~Luhn, S.~Nasri and P.~Ramond,
  %``Tri-bimaximal neutrino mixing and the family symmetry semidirect product of Z(7) and Z(3),''
  Phys.\ Lett.\ B {\bf 652}, 27 (2007)
  doi:10.1016/j.physletb.2007.06.059
  [arXiv:0706.2341 [hep-ph]].
  %%CITATION = doi:10.1016/j.physletb.2007.06.059;%%
  %145 citations counted in INSPIRE as of 09 Oct 2016
  
  %\cite{Kile:2014kya}
\bibitem{Kile:2014kya} 
  J.~Kile, M.~J.~PŽrez, P.~Ramond and J.~Zhang,
  %``$\theta_{13}$ and the flavor ring,''
  Phys.\ Rev.\ D {\bf 90}, no. 1, 013004 (2014)
  doi:10.1103/PhysRevD.90.013004
  [arXiv:1403.6136 [hep-ph]].
  %%CITATION = doi:10.1103/PhysRevD.90.013004;%%
  %7 citations counted in INSPIRE as of 09 Oct 2016
  

  
    %\cite{Vien:2015koa}
\bibitem{Vien:2015koa} 
  V.~V.~Vien,
  %``$T_7$ flavor symmetry scheme for understanding neutrino mass and mixing in 3-3-1 model with neutral leptons,''
  Mod.\ Phys.\ Lett.\ A {\bf 29}, 28 (2014)
  doi:10.1142/S0217732314501399
  [arXiv:1508.02585 [hep-ph]].
  %%CITATION = doi:10.1142/S0217732314501399;%%
  %3 citations counted in INSPIRE as of 08 Feb 2017
  
  %\cite{Vien:2016qbb}
\bibitem{Vien:2016qbb} 
  V.~V.~Vien and H.~N.~Long,
  %``Lepton mass and mixing in a simple extension of the Standard Model based on T7 flavor symmetry,''
  arXiv:1609.03895 [hep-ph].
  %%CITATION = ARXIV:1609.03895;%%
  %1 citations counted in INSPIRE as of 08 Feb 2017
  
    %Delta(27)

  %\cite{Vien:2016tmh}
\bibitem{Vien:2016tmh} 
  V.~V.~Vien, A.~E.~C‡rcamo Hern‡ndez and H.~N.~Long,
  %``The $\Delta(27)$ flavor 3-3-1 model with neutral leptons,''
  Nucl.\ Phys.\ B {\bf 913}, 792 (2016)
  doi:10.1016/j.nuclphysb.2016.10.010
  [arXiv:1601.03300 [hep-ph]].
  %%CITATION = doi:10.1016/j.nuclphysb.2016.10.010;%%
  %14 citations counted in INSPIRE as of 08 Feb 2017
  
  %\cite{Ferreira:2012ri}
\bibitem{Ferreira:2012ri} 
  P.~M.~Ferreira, W.~Grimus, L.~Lavoura and P.~O.~Ludl,
  %``Maximal CP Violation in Lepton Mixing from a Model with Delta(27) flavour Symmetry,''
  JHEP {\bf 1209}, 128 (2012)
  doi:10.1007/JHEP09(2012)128
  [arXiv:1206.7072 [hep-ph]].
  %%CITATION = doi:10.1007/JHEP09(2012)128;%%
  %48 citations counted in INSPIRE as of 08 Feb 2017
  
 
  %PSL(2,7)
  
  %\cite{Chen:2014wiw}
\bibitem{Chen:2014wiw} 
  G.~Chen, M.~J.~PŽrez and P.~Ramond,
  %``Neutrino masses, the $\mu$-term and $\mathcal{ PSL}_2(7)$,''
  Phys.\ Rev.\ D {\bf 92}, no. 7, 076006 (2015)
  doi:10.1103/PhysRevD.92.076006
  [arXiv:1412.6107 [hep-ph]].
  %%CITATION = doi:10.1103/PhysRevD.92.076006;%%
  %5 citations counted in INSPIRE as of 09 Oct 2016
  





 
 
 


  


\end{thebibliography}
\end{document}